%% file: registeredreport.tex
\documentclass[sigconf,authorversion,nonacm]{acmart}

\input{meta/packages.tex}
\input{meta/macros.tex}

\AtBeginDocument{%
  \providecommand\BibTeX{{%
    \normalfont B\kern-0.5em{\scshape i\kern-0.25em b}\kern-0.8em\TeX}}}

\setcopyright{acmcopyright}
\copyrightyear{2022}
\acmYear{2022}
\acmDOI{10.1145/nnnnnnn.nnnnnnn}

\acmConference[MSR 2022]{The 2022 Mining Software Repositories Conference}{May 23--24, 2022}{Pittsburgh, Pennsylvania, USA}
\acmPrice{15.00}

\begin{document}

\title{CamBench - Cryptographic API Misuse Detection Tool Benchmark Suite}
\titlenote{This study was accepted at the MSR 2022 Registered Reports Track.}

\author{Michael Schlichtig}
\authornote{Both authors contributed equally to this research.}
\affiliation{%
	\institution{Heinz Nixdorf Institute at Paderborn University}
	\city{Paderborn}
	\country{Germany}}
\email{michael.schlichtig@uni-paderborn.de}
\orcid{0000-0001-6600-6171}

\author{Anna-Katharina Wickert}
\authornotemark[2]
\affiliation{%
  \institution{Technische Universität Darmstadt}
  \city{Darmstadt}
  \country{Germany}}
\email{wickert@cs.tu-darmstadt.de}
\orcid{0000-0002-1441-2423}

\author{Stefan Krüger}
\affiliation{%
	\institution{Independent}
	\country{Germany}
}
\email{krueger.stefan.research@gmail.com}
\orcid{0000-0003-0895-8830}

\author{Eric Bodden}
\affiliation{%
	\institution{Heinz Nixdorf Institute at Paderborn University \& Fraunhofer IEM}
	\city{Paderborn}
	\country{Germany}}
\email{eric.bodden@uni-paderborn.de}
\orcid{0000-0003-3470-3647}

\author{Mira Mezini}
\affiliation{%
	\institution{Technische Universität Darmstadt}
	\city{Darmstadt}
	\country{Germany}}
\email{mezini@cs.tu-darmstadt.de}
\orcid{0000-0001-6563-7537}

\renewcommand{\shortauthors}{Schlichtig and Wickert, et al.}

\begin{abstract}
	\input{sections/00_abstract.tex}
\end{abstract}


\maketitle

\input{sections/10_introduction.tex}

\input{sections/20_background.tex}

\input{sections/30_researchQuestions.tex}

\input{sections/40_Methodology.tex}
\input{sections/50_executionPlan.tex}

\input{sections/70_validity.tex}
\input{sections/60_Implications.tex}

\begin{acks}
	Funded by the Deutsche Forschungsgemeinschaft (DFG, German Research Foundation) – SFB 1119 – 236615297 and by the German Federal Ministry of Education and Research and the Hessen State Ministry for Higher Education, Research and the Arts within their joint support of the National Research Center for Applied Cybersecurity ATHENE. 
\end{acks}

\bibliographystyle{ACM-Reference-Format}
\bibliography{bibliography.bib}

\end{document}

%% file: meta/packages.tex
\usepackage[utf8]{inputenc}
\usepackage{algorithmic}
\usepackage{graphicx}
\usepackage{textcomp}
\usepackage{xcolor}
\usepackage{filecontents} 
\def\BibTeX{{\rm B\kern-.05em{\sc i\kern-.025em b}\kern-.08em
    T\kern-.1667em\lower.7ex\hbox{E}\kern-.125emX}}
\usepackage[textsize=tiny, color=green!50]{todonotes}
\usepackage[nolist]{acronym} %
\usepackage{tikz} %
\usetikzlibrary{positioning} %
\usepackage{booktabs}
\usepackage{colortbl}
\usepackage{listings}
\usepackage{verbatim}
\usepackage{url}
\usepackage{meta/lstcustom}
\usepackage{caption}
\usepackage{subcaption}
\usepackage{csvsimple} %
\usepackage[english]{babel}
\usepackage[bottom]{footmisc}
\usepackage{hyperref}
\usepackage{flushend}
\usepackage{orcidlink}
\usepackage{stfloats}

%% file: meta/macros.tex
\newcommand{\checkNum}[1]{{\textcolor{black}{\textit{#1}}}}
\newcommand{\checknum}[1]{\checkNum{#1}}
\newcommand{\authorResponse}[1]{#1}
\newcommand{\topicsentence}[1]{#1}

\definecolor{gray}{gray}{0.85}
\definecolor{light-gray}{gray}{0.9}
\definecolor{lgray}{gray}{0.9}
\definecolor{firebrick}{rgb}{0.7, 0.13, 0.13}
\definecolor{mountainmeadow}{rgb}{0.19, 0.73, 0.56}

\definecolor{carmine}{rgb}{0.59,0.0,0.09}
\definecolor{light-gray}{gray}{0.80}

\newcommand{\cryptobench}{\textit{CryptoAPI-Bench}}
\newcommand{\owaspbench}{\textit{OWASP Benchmark}}
\newcommand{\bragabench}{\textit{Braga Benchmark}}
\newcommand{\mubench}{\textit{MuBench}}
\newcommand{\mubenchparamcrypto}{\textit{Parametric Crypto Misuse Benchmark}}
\newcommand{\ghera}{\textit{Ghera Benchmark}}
\newcommand{\cc}{\textit{CogniCrypt\textsubscript{SAST}}}
\newcommand{\cognicrypt}{\cc{}}

\newcommand{\cg}{\textit{CryptoGuard}}
\newcommand{\cryptoguard}{\cg{}}

\newcommand{\spotbugs}{\textit{SpotBugs}}
\newcommand{\coverity}{\textit{Coverity}}
\newcommand{\findsecbugs}{\textit{FindSecBugs}}
\newcommand{\visualcodegrepper}{\textit{VisualCodeGrepper}}
\newcommand{\xanitizer}{\textit{Xanitizer}}
\newcommand{\sonarqube}{\textit{SonarQube}}
\newcommand{\yasca}{\textit{Yasca}}
\newcommand{\amandroid}{\textit{Amandroid}}
\newcommand{\appcritique}{\textit{AppCritique}}
\newcommand{\devknox}{\textit{DevKnox}}
\newcommand{\fixdroid}{\textit{FixDroid}}
\newcommand{\marvinsa}{\textit{Marvin-SA}}
\newcommand{\mobsf}{\textit{MobSF}}
\newcommand{\julia}{\textit{Julia}}
\newcommand{\owaspzap}{\textit{OWASP ZAP}}
\newcommand{\pmd}{\textit{PMD}}
\newcommand{\obench}{\textit{CamBench}}

\newcommand{\Obench}{\obench{}}

\newcommand{\dacapo}{\textit{DaCapo benchmarks}}
\newcommand{\abm}{\textit{ABM}}
\newcommand{\boa}{\textit{BOA}}
\newcommand{\qualitas}{\textit{Qualitas Corpus}}
\newcommand{\hermes}{\textit{Hermes}}

\newcommand*{\etal}{et~al.}

\makeatletter %
\newcommand{\linebreakand}{%
\end{@IEEEauthorhalign}
\hfill\mbox{}\par
\mbox{}\hfill\begin{@IEEEauthorhalign}
}
\makeatother %

\newcommand{\filleddot}[1]{%
	$%
	\vcenter{%
		\hbox{%
			\begin{tikzpicture}%
			\draw[fill={rgb:red,50;green,153;yellow,0}, draw={rgb:red,50;green,153;yellow,0}] (0,0) -- (0:#1) arc (0:360:#1) -- cycle;
			\end{tikzpicture}%
		}%
	}%
	$%
}

\newcommand{\emptydot}[1]{%
	$%
	\vcenter{%
		\hbox{%
			\begin{tikzpicture}%
			\draw[draw={rgb:red,50;green,0;yellow,0}, thin] (0,0) circle (#1);
			\end{tikzpicture}%
		}%
	}%
	$%
}

%% file: sections/00_abstract.tex
\textbf{Context:} Cryptographic APIs are often misused in real-world applications.
To mitigate that, many cryptographic API misuse detection tools have been introduced.
However, there exists no established reference benchmark for a fair and comprehensive comparison and evaluation of these tools.
While there are benchmarks, they often only address a subset of the domain or were only used to evaluate a subset of existing misuse detection tools.
\textbf{Objective:} To fairly compare cryptographic API misuse detection tools and to drive future development in this domain, we will devise such a benchmark.
Openness and transparency in the generation process are key factors to fairly generate and establish the needed benchmark.
\textbf{Method:} We propose an approach where we derive the benchmark generation methodology from the literature which consists of general best practices in benchmarking and domain-specific benchmark generation.
A part of this methodology is transparency and openness of the generation process, which is achieved by pre-registering this work.
Based on our methodology we design \obench, a fair \grqq\textbf{C}ryptographic \textbf{A}PI \textbf{M}isuse Detection Tool \textbf{Bench}mark Suite\grqq{}.
We will implement the first version of \obench\ limiting the domain to Java, the JCA, and static analyses.
Finally, we will use \obench\ to compare current misuse detection tools and compare \obench\ to related benchmarks of its domain.

%% file: sections/10_introduction.tex
\section{Introduction}
\label{sec:intro}

Cryptography, hereafter crypto, is widely used in today's software to ensure the confidentiality of users' data. 
Concretely, through crypto, everyone can securely use online banking, purchase a book from the local book shop from their computer, and share sensitive information with their co-workers while working remotely without the fear that the software leaks their data. 
Unfortunately, previous research results show that developers struggle with the secure usage of crypto APIs and often add vulnerabilities to their code~\cite{nadi2016jumping,gorski2018developers,lazar2014does,wickert2021python}. 
Most of these vulnerabilities are caused by developers who use a crypto API in a way that is considered insecure by experts, e.g., choosing an outdated crypto hash algorithm like \textit{SHA-1}~\cite{lazar2014does}.
We use the term crypto-API misuse or shortly misuse to describe these programming flaws. 
Further, we focus on Java crypto-APIs to provide concrete results.
To support developers and security auditors to identify these misuses, many crypto API misuse detectors such as \textit{CryptoRex}~\cite{zhang2019cryptorex}, \textit{CryptoLin}t~\cite{egele2013empirical}, \cognicrypt~\cite{kruger2019crysl}, and \cryptoguard~\cite{rahaman2019cryptoguard} have been developed.

While several tools demonstrate the practical importance and their capabilities to identify issues in real-world code, a fair comparison of the analysis capabilities is difficult. 
First, all in-the-wild studies were conducted on different applications and domains. 
For example, \textit{CryptoLint}~\cite{egele2013empirical} was evaluated on \checkNum{11,748} Android apps in \checkNum{2012} while \cognicrypt~\cite{kruger2019crysl} analyzed \checknum{10,000} apps in \checkNum{2017}. 
While both tools demonstrated their capabilities for Android apps, it is hard to judge how they compare, as they analyzed different apps and the apps evolved over time. 
Further, other domains of Java software are only analyzed by one tool, e.g., Apache applications, as representatives for large Java projects~\cite{rahaman2019cryptoguard}, or Maven artifacts, representatives for Java libraries~\cite{kruger2019crysl}.
Thus, a fair comparison of these analyses and their capabilities is challenging. 

An accepted standard in other fields to overcome the above-described problem are benchmarks. 
While it is time-consuming to create a proper benchmark, the advantages are significant.
A benchmark can set a common understanding within the community of the capabilities of the respective solutions, e.g., the strength of a misuse detector. 
If a benchmark is used for several tools, which should be the aim, a benchmark enables a fair comparison of the respective solutions in the field. 
Further, a proper benchmark can thrive improvements of each tool and form a community to improve the state of the art significantly. 
Overall, an established benchmark drives the development and research in the community of its domain~\cite{DaCapo06}.
 
Unfortunately, for the domain of crypto-API misuse detection, one still misses a standard benchmark. 
While we, as a community, have a few custom benchmarks~\cite{afrose2019cryptoapi,braga2017practical,wickert2019dataset,mitra2017ghera}, all of them have significant limitations. 
Some benchmarks are developed for a specific problem, e.g., parametric-based misuses~\cite{wickert2019dataset}, and are not suitable outside of this problem space. 
Other benchmarks come with a significant bias as their creation was along with the development of the respective detection tool~\cite{afrose2019cryptoapi}.

To solve the above-described problem, we aim to establish \textbf{\obench}, a fair \grqq\textbf{C}ryptographic \textbf{A}PI \textbf{M}isuse Detection Tool \textbf{Bench}mark Suite\grqq{} to compare crypto-API misuse detectors.
Thus, we will analyze the current state of the art of benchmarking crypto-API misuse detection tools, and how these benchmarks adhere to proper methods for benchmarking.  
Further, we plan to focus on understanding and deriving specific challenges that the domain of crypto API misuse detection tools poses to the design of benchmarks, e.g., the evolution of security characteristics. 
Based on this theoretical work, we want to create our fair benchmark to compare crypto-API misuse detectors.

Our benchmark suite consists of \checknum{two} benchmarks and \checknum{one} heuristic to compare crypto-API misuse detectors. 
The benchmarks are a collection of real-world applications as well as synthetic test cases to evaluate the analysis capabilities. 
In addition to the benchmarks, we add an API coverage heuristic that will provide detailed information about the strengths and limitations of a misuse detector with respect to certain API classes. 
This property is important as an analysis which misses a class may miss a severe vulnerability in an application due to the respective class. 
Thus, an assesment of this property beside precision and recall is important. 
To understand the impact of creating an independent benchmark, we will compare \obench\ against the existing benchmarks. %

The first version of \obench\ focuses on usages of the standard crypto library in Java, the Java Cryptography Architecture (JCA), and static analyses. 
To support future extensions, e.g., on different APIs and dynamic analyses, one design principle of the benchmark is extensibility. 
We plan to open-source all our research artifacts along with our tooling.
This way, we enable the extension of \obench\ and provide tooling to simplify the creation of future benchmarks in related domains.  
Further, our tooling and our theoretical work can further reduce the effort to create fair benchmarks for arbitrary APIs, e.g., misuse of the \textit{Iterator} API.

The early validation of the design of \obench\ and the benchmark generation methodology is crucial to achieving the aim of a fair crypto-API misuse detector benchmark because it aides in making the process more transparent and open as it also allows for community feedback.
Further, the validation is of importance to avoid tuning of \obench\ in a specific direction by the authors, as well as to avoid requests for changes due to the results by a reviewer who may be an author of one of the benchmarked tools. 
Moreover, \obench\ is created by members of the community and as part of the community, we value this effort. 
Therefore, we believe that the pre-registration will strengthen the validity of \obench\, the results we derive, and reduces bias in the creation.

%% file: sections/20_background.tex
\input{tables/benchmarksToSA.tex}

\section{Preliminary Study: Benchmarks Covering Cryptography}
\label{sec:background}

To understand the differences, strengths, and weaknesses of the existing benchmarks for crypto misuse detectors, we will present our preliminary study on the current state of the art of benchmarks for crypto misuse detection tools. 
We derived all discussed benchmarks by collecting all existing crypto benchmarks we were aware of and extended them through a forward- and backward search of the related publications. 
An overview of all identified benchmarks and misuse detectors that were successfully evaluated with them is presented in Table~\ref{tab:benchToSA}.
Note, that we did not list any analysis for \mubench\ as the API misuse detection tools used for evaluation did not find any crypto misuse~\cite{amann2018systematic}. 
Similarly, we also omit the misuse detection tools that mark the secure and insecure test cases of the \ghera\ as non-security critical (cf. Ranganath \etal~\cite{ranganath2020free}) for readability. %

\topicsentence{\cryptobench~\cite{afrose2019cryptoapi} is the most-recent benchmark and consists of synthetic examples to identify misuses.}
In total, \checknum{171} examples were created along \checknum{16} different vulnerabilities for \checknum{basic} (40) and \checknum{advanced} (131) cases for static analyses.
The advanced tests focus on inter-procedurality, field-sensitivity, and path-sensitivity. 
Thus, challenging the static analysis for several different analysis capabilities. 
However, the benchmark comprises \checknum{all} the vulnerabilities detected by \cryptoguard~\cite{rahaman2019cryptoguard}. 

\topicsentence{While \cryptobench~focuses on analysis capabilities, the \bragabench~\cite{braga2017practical} focuses on providing developers with the best analysis tool for their team.}
A team is characterized by their experience (novice or knowledgeable) and their support for cryptographic tasks (unsupported or supported). 
In addition to the team's characteristics, the test cases of the benchmark are grouped along \checknum{three} complexity classes of crypto misuses. 
The complexity is represented as the abstraction level required to identify and respectively fix the misuse. 
For example, identifying an outdated hashing algorithm is considered as low complexity, while the reuse of a nonce is marked as a high complexity problem caused by an insecure system design or architecture. 
Thus, this benchmark sheds light on different complexities to cover different teams' knowledge. 

\topicsentence{In contrast to the previous two benchmarks, the ~\mubenchparamcrypto~\cite{wickert2019dataset} derived misuse cases from in-the-wild code.}
All misuses were collected from top Java GitHub projects and cover crypto misuses caused by passing an insecure parameter to a function. 
Thus, covering one of the major problems, incorrect configurations via parameters, observed in several in-the-wild studies~\cite{kruger2019crysl,egele2013empirical} while ignoring more complex misuses, e.g., incorrect call order or misuses caused by architecture flaws. 
While the previously discussed benchmarks created the benchmarking infrastructure, \mubenchparamcrypto\ extended an existing and well established benchmark for API-misuses~\cite{amann2016mubench}.

\topicsentence{The \mubenchparamcrypto~benchmark is built upon \mubench~\cite{amann2016mubench} which is a benchmark for general API misuses, including several crypto misuses in Java.}
In the publication from \checknum{2016}, the benchmark includes \checkNum{18} misuses manually identified from bug fixes on GitHub and SourceForge. 
All of these cases, except \checknum{one} fix, influence the behavior of the program, such as a crash due to invalid input. 
Thus, the misuses cover spurios behavior rather than insecure algorithm choices.  
Later, the benchmark was extended with additional \checknum{31} misuses by mining projects via \boa~\cite{dyer2013boa} which use the \textit{javax.crypto.Cipher} or the entire \textit{javax.crypto} package, and manually verifying the security of the code~\cite{amann2018systematic}.

\topicsentence{The \ghera~\cite{mitra2017ghera} focuses on general vulnerabilities in Android applications being a subset of Java misuses, and thus cover a few crypto misuse cases as well.}
In total, this benchmark includes \checknum{five} instances of crypto vulnerabilities and \checknum{eight} vulnerabilities caused by networking. 
While it may be counter-intuitive, the later \checknum{eight} vulnerabilities may be covered by crypto misuse detectors as well when covering SSL and TLS related misuses, e.g., the usage of TLS 1.0. 
While the number is low, all cases are executable Android applications with a vulnerability along with an exploit of the respective vulnerability. 

\topicsentence{While the previous benchmarks are developed in academia, the \owaspbench~\cite{wichers_owasp_nodate} is developed by industry. }
Similar to our motivation, the underlying motivation for \owaspbench\ was to measure the strength and weaknesses of different static analyses to detect vulnerabilities such as crypto misuses. 
To achieve this aim, version \checknum{1.2} consists, similarly to \ghera, of test cases with real-world web applications that can be exploited. 
The exploits are categorized along with the different vulnerability types of OWASP Top-10 web vulnerabilities~\cite{vanderstock_owasptop10_2021}, such as command injection and weak cryptography.
In total, \checknum{246} tests for vulnerabilities are in \owaspbench. 

\topicsentence{The overlap of the static analyses, e.g., \cognicrypt~\cite{kruger2019crysl}, \cryptoguard~\cite{rahaman2019cryptoguard}, and \findsecbugs~\cite{arteau_find_2020}, analyzed on the aforementioned benchmarks is rather small as the motivation and domain for which all these benchmarks were created differs.}  
We illustrate the benchmarked static analyses for each previously discussed benchmark in Table~\ref{tab:benchToSA}.
Only \checknum{two} tools, namely \findsecbugs~(three) and \sonarqube~(two), are evaluated on more than one benchmark.
The remaining \checkNum{16} analyses are evaluated only on one benchmark.  
For example, the \ghera~evaluated analyses for Android vulnerabilitites, such as \checkNum{\amandroid~\cite{wei2014amandroid}}.
On the other hand, tools focused on crypto misuses, such as \cryptoguard~\cite{rahaman2019cryptoguard}, are evaluated on the crypto-specific \cryptobench.

%% file: tables/benchmarksToSA.tex
\begin{table*}[b]
    \centering
    \begin{tabular}{p{3.2cm}llllllllllllllllllll}
    \toprule
    Benchmark & A & AC & CC & CG & CO & DN & FD & FS & J & MA & MF & P & SB & SQ & V & X & Y & Z & Source \\
    \midrule
    \cryptobench~\cite{afrose2019cryptoapi} & \emptydot{0.1} & \emptydot{0.1} & \filleddot{0.1} & \filleddot{0.1} & \filleddot{0.1} & \emptydot{0.1} & \emptydot{0.1} & \emptydot{0.1} & \emptydot{0.1} & \emptydot{0.1} & \emptydot{0.1} & \emptydot{0.1} & \filleddot{0.1} & \emptydot{0.1} & \emptydot{0.1} & \emptydot{0.1} & \emptydot{0.1} & \emptydot{0.1} & \cite{afrose2019cryptoapi} \\ 
    \bragabench~\cite{braga2017practical} & \emptydot{0.1} & \emptydot{0.1}\ & \emptydot{0.1} & \emptydot{0.1} & \emptydot{0.1} & \emptydot{0.1} & \emptydot{0.1} & \filleddot{0.1} & \emptydot{0.1} & \emptydot{0.1} & \emptydot{0.1} & \emptydot{0.1} & \emptydot{0.1} & \filleddot{0.1} & \filleddot{0.1} & \filleddot{0.1} & \filleddot{0.1} & \emptydot{0.1} & \cite{braga2017practical} \\ 
    \mubenchparamcrypto~\cite{wickert2019dataset} & \emptydot{0.1} & \emptydot{0.1} & \emptydot{0.1} & \emptydot{0.1} & \emptydot{0.1} & \emptydot{0.1} & \emptydot{0.1} & \filleddot{0.1} & \emptydot{0.1} & \emptydot{0.1} & \emptydot{0.1} & \emptydot{0.1} & \emptydot{0.1} & \emptydot{0.1} & \emptydot{0.1} & \emptydot{0.1} & \emptydot{0.1} & \emptydot{0.1} & \cite{wickert2019dataset} \\ 
    \mubench~\cite{amann2016mubench} & \emptydot{0.1} & \emptydot{0.1} & \emptydot{0.1} & \emptydot{0.1} & \emptydot{0.1} & \emptydot{0.1} & \emptydot{0.1} & \emptydot{0.1} & \emptydot{0.1} & \emptydot{0.1} & \emptydot{0.1} & \emptydot{0.1} & \emptydot{0.1} & \emptydot{0.1} & \emptydot{0.1} & \emptydot{0.1} & \emptydot{0.1} & \emptydot{0.1} & \cite{amann2016mubench} \\ 
    \ghera~\cite{mitra2017ghera} & \filleddot{0.1} & \filleddot{0.1} & \emptydot{0.1} & \emptydot{0.1} & \emptydot{0.1} & \filleddot{0.1} & \filleddot{0.1} & \emptydot{0.1} & \emptydot{0.1} & \filleddot{0.1} & \filleddot{0.1} & \emptydot{0.1} & \emptydot{0.1} & \emptydot{0.1} & \emptydot{0.1} & \emptydot{0.1} & \emptydot{0.1} & \emptydot{0.1} & \cite{ranganath2020free} \\ 
    \owaspbench~\cite{wichers_owasp_nodate} & \emptydot{0.1} & \emptydot{0.1} & \emptydot{0.1} & \emptydot{0.1} & \emptydot{0.1} & \emptydot{0.1} & \emptydot{0.1} & \filleddot{0.1} & \filleddot{0.1} & \emptydot{0.1} & \emptydot{0.1} & \filleddot{0.1} & \emptydot{0.1} & \filleddot{0.1} & \emptydot{0.1} & \emptydot{0.1} & \emptydot{0.1} & \filleddot{0.1} & \cite{burato2017security} \\ 
    \bottomrule
    \end{tabular}
    \caption{An overview of all identified benchmarks and the crypto misuse detectors \small{
		("*" marks non-academic static analyses.
	     	 A: \amandroid~\cite{wei2014amandroid}, 
      	        AC: \appcritique *~\cite{appcritique},
	     	CC: \cognicrypt~\cite{kruger2019crysl},
	     	CG: \cryptoguard~\cite{rahaman2019cryptoguard},
      	        CO: \coverity *~\cite{synopsys_inc_coverity_nodate},
    	 	DN: \devknox *~\cite{xysec_labs_devknox_nodate},
		FD: \fixdroid~\cite{nguyen2017stitch},
		FS: \findsecbugs *~\cite{arteau_find_2020},
		J: \julia *~\cite{spoto2016julia},
		MA: \marvinsa *~\cite{rinaudo_marvin_2022},
		MF: \mobsf *~\cite{abraham_mobile_2022},
		P: \pmd *~\cite{noauthor_pmd_2022},
	     	SB: \spotbugs *~\cite{noauthor_spotbugsspotbugs_2022}, 
		SQ: \sonarqube *~\cite{sonarsource_sonarqube_2022},
		V: \visualcodegrepper *~\cite{noauthor_visualcodegrepper_2022}, 
		X: \xanitizer *~\cite{rigs_it_gmbh_xanitizer_nodate},  
		Y: \yasca *~\cite{scovetta_yasca_2022}, and 
		Z: \owaspzap *~\cite{noauthor_owaspzap_2022}
		)} analyzed. 
	    }
    \label{tab:benchToSA}
\end{table*}

%% file: sections/30_researchQuestions.tex
\section{Research Questions}
\label{sec:researchQuestions}
\topicsentence{With the research proposed in this registered report, we will answer the \checkNum{three} research questions discussed in this Section.}
We start with a question that we already discussed in Section~\ref{sec:background}, and plan to investigate more in-depth:
\paragraph{RQ1}\label{RQ_SLR} \textit{What is the current state of the art of benchmarking crypto API misuse detection tools and to which extent do benchmarks adhere to appropriate methods in general benchmarking?}
As we have shown in Section~\ref{sec:background}, there is a need for a fair benchmark that properly covers the domain of crypto API misuse detection tools in an unbiased way.
Existing benchmarks are not suitable for a fair comparison as analysis properties, such as sensitivity, are not considered or not covered completely, or covered misuses are designed alongside one existing tool rather than the analyzed crypto APIs. 

Since existing benchmarks are insufficient, a proper generation of a new benchmark is necessary.
Key factors for this are transparency and openness, as the successful DaCapo benchmark project has shown~\cite{DaCapo06}.
To achieve this goal, performing the benchmark generation via a registered report is a logical consequence, as the proposed methodology is reviewed by experts upfront and the process is made known to the research community.

\paragraph{RQ2}\label{RQ_methodology} \textit{In the area of evaluating cryptographic API misuse detection tools, what requirements does a benchmark need to be tailored to?}

For the generation of a fair benchmark, we first need to define our methodology, which we derive from the literature.
Understanding the domain-specific characteristics is necessary to develop the requirements a benchmark for evaluation crypto misuse detection tools needs to meet to support their evaluation and comparison.
Once the methodology and requirements are defined, the next step is generating \obench\ accordingly and then evaluating it.

\paragraph{RQ3}\label{RQ_Evaluation} \textit{Can \obench\ meaningfully compare current cryptographic API misuse detection tools?}
Our final research question is the application of \obench\ on static analysis tools for the misuse detection of the JCA in Java code.
We plan to evaluate two aspects.
First, we will use \obench\ to compare how current crypto API misuse detection tools perform.
Second, we will compare \obench\ to the other benchmarks in the domain of crypto API misuse detection tools (cf. Table~\ref{tab:benchToSA}) and analyze the differences.

%% file: sections/40_Methodology.tex
\section{Methodology}
\label{sec:methodology}
In this section, we discuss the theoretical foundations of our benchmark generation methodology and design.
We have already begun conducting a literature survey to answer \hyperref[RQ_SLR]{\textbf{RQ1}}, the preliminary results of which we presented in Section~\ref{sec:background}.
As of now, our results indicate that there is no widely accepted or established benchmark for the evaluation of crypto API misuse detection tools.
Existing benchmarks have only been used to compare subsets of crypto misuse detection tools (cf.~Table~\ref{tab:benchToSA}).
To fully answer this question, we will continue our literature review and supplement our findings. For this purpose, we will adapt the procedure described by Kitchenham \etal~\cite{kitchenham2004procedures}.

\topicsentence{To answer \hyperref[RQ_methodology]{\textbf{RQ2}} we consider two aspects:}
First, we discuss what the proper methodology for benchmark generation in general is, and how this can be applied to one domain, namely crypto API misuse detection tool evaluation in Section~\ref{subsec:literatureBenchmarks}.
And second, we analyze the special characteristics of this domain that need to be addressed by \obench\ or in general a benchmark for this domain in Section~\ref{subsec:domainAnalysis}.
We conclude the theoretical foundations by deriving the domain-specific requirements for \obench\ and  developing the domain-specific methodology for the generation of benchmarks in Section~\ref{subsec:requirementsBenchmark}.
Then we use both to formulate the practical application in our execution plan for the generation of \obench\ and its evaluation to answer \hyperref[RQ_Evaluation]{\textbf{RQ3}} in Section~\ref{sec:executionPlan}. %

\subsection{Literature review on benchmark generation approaches}\label{subsec:literatureBenchmarks}
\topicsentence{The creation of a benchmark for a specific domain requires first an understanding of the characteristics of well-established more general benchmarks in the community.}
So far, our literature review to establish a benchmark for crypto API misuse detection tools has yielded a few relevant approaches of more general benchmarks.
For our approach two findings are standing out, namely the \dacapo~\cite{DaCapo06} and the Automated Benchmark Management System (\abm)~\cite{ABM16}.
The \dacapo\ are a collection of widely used open-sourced benchmarks for the Java programming language introduced in 2006 and its latest release in 2018.
In their report on the development and establishment of the \dacapo, Blackburn et al. stressed the importance of an open and transparent process with community feedback.
Their success is a compelling argument to follow transparency and community feedback as guiding principles when developing \obench.
Similarly, the \qualitas~\cite{tempero2010qualitas} also provides a large collection of real-world open-source Java code. %
The way the real-world applications are organized and enriched with metadata, also for versioning, is valuable information for the design of benchmarks regarding extensibility.
However, the \qualitas\ was introduced in 2010 and curated, but with its latest update in 2013 it appears to be unmaintained.

The \dacapo\ and \qualitas\ are both benchmarks for Java in general targeting the evaluation of the performance of Java code and Java virtual machines.
However, evaluating tools that detect misuses of crypto APIs is a different task.
The main focus here rather lies on the correctness of the results, i.e., precision and recall of the detected issues and the soundness of the analysis itself.
Moreover, how usable such a crypto misuse detection tool is, also relies on its precision and recall~\cite{christakis2016developers} -- too many false positives are one of the main reasons why developers omit using static analysis tools. %
Such domain-specific requirements can be better addressed by the approach of Nguyen Quang Do et al. called \abm~\cite{ABM16} since domain-specific data sets are needed to evaluate specific research questions.
They describe how to semi-automatically generate and maintain domain-specific benchmark suites.
The benchmark suites are collected with a well-described process of crawling buildable open-source projects of the target domain.
Therefore, they provide a representative set of real-world applications relevant in the domain. 
We derive our methodology from the experience that can be taken from the \dacapo\ and the proposed procedure by \abm\ and provide a more in-depth explanation of the execution in Section~\ref{sec:executionPlan}.
Moreover, \abm\ already describes that filtering the real-world applications for relevance is integral to being representative.
Hence, understanding the requirements of a domain is key to generating a representative benchmark suite for it.

\subsection{Analysis of the domain for benchmarking cryptographic API misuse detection tools}\label{subsec:domainAnalysis}
\topicsentence{While Section~\ref{sec:background} presents preliminary findings of benchmarks targeted for crypto API misuse detectors, a structured literature review~\cite{kitchenham2004procedures} is required to enrich our findings and ensure their completeness.}
The goal that crypto API misuse detection tools aim to achieve is detecting and reporting crypto misuses in applications that use crypto APIs.
Employing crypto in general sets the focus on security.
Therefore, when comparing misuse detection tools it is important to evaluate their performance in terms of crypto API coverage, precision and recall, and analysis capabilities.
The coverage is important for developers to know whether their code using a crypto API is completely or only partially covered, e.g., common known vulnerabilities.
Research on reasons why misuse detection tools are rarely used has shown that precision and recall of a tool play a key role~\cite{christakis2016developers} -- too many false positives are a major reason for developers to omit using a tool~\cite{Johnson2013}.
Moreover, the analysis evaluation of analysis capabilities is important as well.
Whether an analysis is intraprocedural, inter-procedural, flow-sensitive, context-sensitive, field-sensitive, path-sensitive, and object-sensitive or not has direct implications on what the analysis is actually able to detect.
Unfortunately, the benchmarks identified in Section~\ref{sec:background} either ignore such properties or only include a subset, such as \cryptobench~\cite{afrose2019cryptoapi}.

\subsection{Domain-specific Benchmark Requirements}\label{subsec:requirementsBenchmark}
\topicsentence{Concluding the theoretical foundations of our methodology, we specify the requirements specific to the domain of benchmarking crypto API misuse detection tools and propose our benchmark design besaed on the results of the previous sections.}
Following the take-aways from the \dacapo~\cite{DaCapo06} (transparent and open process, a variety of real-world applications maximizing the domain coverage, easy to use, and providing metrics), \obench\ will be a benchmark suite consisting of several benchmarks.
A good example for easy to use is the well usable deployment system of \emph{MuBench}~\cite{amann2016mubench}, which delivers its execution pipeline as a Docker container and provides extensive documentation on how to use and adapt the benchmark as well as on how to contribute.
Moreover, we enrich the collection process of real-world application with the semi-automated approach of \abm~\cite{ABM16}.

Based on these presented approaches, we derive the following requirements that \obench\ must meet:
\begin{enumerate}
	\item The benchmark generation process needs to be transparent and consider community feedback.
	\item The benchmark contains a representative and diverse collection of real-world applications using crypto APIs. Projects fulfill these properties: \emph{open-source}, \emph{compilable (source code and binaries are available)}, and \emph{representative of the domain's real-world application, e.g., diverse size, context, $\dots$}.
	\item The benchmark is open-sourced.
	\item The benchmark is provided with a deployment system.
\end{enumerate}

\obench\ will be a suite of benchmarks covering the following three main aspects: \textit{(a)} \emph{real-world applications with crypto API usage} to evaluate the performance of misuse detection tools on relevant applications~\cite{ABM16}, \textit{(b)} \emph{analysis capabilities}, and \textit{(c)} \emph{crypto API coverage} to evaluate which fraction of the API a misuse detection tool supports.
We added \emph{analysis capabilities} and  \emph{crypto API coverage} to accommodate the requirements of the domain of benchmarking crypto API misuse detection tools.

\topicsentence{To facilitate the evaluation of analysis capabilities, we will develop synthetic test cases for all relevant properties.}
When building an analysis tool, it is necessary to make assumptions that can lead to unsoundness~\cite{reps_Undecidability, livshits2015defense}.
Therefore, evaluating the analysis capabilities which are part of such assumptions is important to draw implications concerning a tool's unsoundness and precision.
The properties we plan to include in \obench\ are flow-sensitivity, context-sensitivity, field-sensitivity, object-sensitivity, and path-sensitivity.

\topicsentence{\Obench\ will provide a benchmark for testing the coverage of Java crypto-APIs.}
Similar to \abm's~\cite{ABM16} focus on automation our goal is to develop a process with a high degree of automation for the generation of the benchmarks for the coverage of the JCA. 
This will help in maintaining and updating the benchmark in the future.

Furthermore, we will enrich the test cases with  metadata similar to \emph{MuBench}~\cite{amann2016mubench} where misuse code examples are specified with YAML files enriched by additional information like misuse type and description.
They also provide examples for correct usages.
Moreover, the metadata comprises information whether a test case is a (in-)secure crypto usage (whether a crypto usage is (in-)secure might also be context-dependent, e.g., using a hash algorithm like MD5 is considered insecure~\cite{kasgar2013review}, yet using MD5 in the context of file validation~\cite{AWS} might still be acceptable), usage category~\cite{amann2018systematic}, and the severity.
Moreover, we will consider metadata on versioning of real-world projects as proposed in the \qualitas~\cite{tempero2010qualitas}.

\topicsentence{To summarize, \obench\ will provide a collection of \emph{real-world applications} with crypto API usage following \abm~\cite{ABM16}, synthetic tests to evaluate the crypto API misuse detectors \emph{analysis capabilities}, and a collection of test cases for checking the \emph{crypto API coverage} of the JCA.}%

%% file: sections/50_executionPlan.tex
\section{Execution Plan}
\label{sec:executionPlan}
In this section, we first describe the generation of \obench\ and its three kinds of benchmarks in Section~\ref{subsec:generationBenchmark}, namely \emph{real-world applications}, \emph{analysis capabilities}, and \emph{crypto API coverage}.
Then we will discuss the publishing and deployment process of \obench\ in Section~\ref{subsec:publishingDeployment}.
Finally, we describe our intended evaluation in Section~\ref{subsec:evelauation}.

\subsection{Generation Plan}\label{subsec:generationBenchmark}
\topicsentence{For the generation of \obench's three benchmarks, we now describe our plan to execute our methodology (cf. Section~\ref{sec:methodology}).}
For the concrete implementation of our methodology for the first version of \obench, we set the scope to the standard crypto library in Java, the JCA, and static analysis tools detecting misuses of it.
\paragraph{Benchmark for real-world applications}\label{par:realworldAPPs}
\topicsentence{For the collection and creation of the real-world application benchmark, we adapt the process specified by \abm~\cite{ABM16}.}
This process consists of six steps of collecting, filtering, and building open-source projects.
The first three steps are \textit{(i)} \emph{mining open-source projects}, \textit{(ii)} filtering for \emph{active projects} and \textit{(iii)} filtering for \emph{known build systems}.
These steps can be automated.
The subsequent steps require manual effort.
In their instantiation of \abm\ for the domain of Java business web applications, Nguyen Quang Do et al. chose GitHub as a platform for mining open-source projects~\cite{ABM16}.
Since this has worked successfully and GitHub is a popular and well-used platform for hosting open-source projects, we intend to use it, too.
The next step is \textit{(iv)} downloading and \emph{building the projects}.
The real-world application benchmark only includes projects that build with standard build systems. 
Another filtering step follows where, in our case, we will \textit{(v)} check whether the \emph{JCA is used} in the projects.
The last step is \textit{(iv)} \emph{compiling the binaries} and adding projects with binaries to the benchmark dataset (cf. requirements in Section~\ref{sec:methodology}).

Furthermore, we add a step to the approach of \abm~\cite{ABM16} by adding metadata files including information, on whether the usage is (in-)secure for selected usages of the JCA.
\authorResponse{To automatically identify usages of the JCA in code, we plan to employ static analyses, such as adapted versions of CogniCrypt~\cite{kruger2019crysl} or CryptoGuard~\cite{rahaman2019cryptoguard}.}
For the ground-truth, whether a JCA usage is secure or insecure, we will use the guidelines of NIST~\cite{NIST_B}, SOG-IS~\cite{SOGIS}, or BSI~\cite{BSI-TR02102} \authorResponse{for manual labeling}.
Since the guidelines of the BSI are updated regularly and are the ones updated most recently, we will primarly be using them.
Furthermore, the employed guidelines can be documented in the metadata files.
The design of the metadata files will be derived from \mubench~\cite{amann2016mubench} and take the experiences concerning versioning from \textit{Qualitas Corpus}~\cite{tempero2010qualitas} into account.
The expected dataset of the real-world application will contain too many usages of the JCA for the manual creation of the metadata files.
Therefore, we will select a representative subset of the JCA usages to write the metadata files.

\paragraph{Benchmark for analysis capabilities}
\topicsentence{The benchmark for evaluating analysis capabilities will consist of synthetic test cases specifically designed to test relevant properties.}
The test cases for the analysis capabilities require manual effort of domain experts. 
They will include test cases for flow-sensitivity, context-sensitivity, field-sensitivity, object-sensitivity, and path-sensitivity, as well as intraprocedural and inter-procedural test cases.
The selected sensitivities are accepted as relevant capabilities for static analyses~\cite{Li2017static,qiu2018analyzing}.
We will develop all test cases for this benchmark and also provide the test cases with the metadata files for (mis-)uses.

\paragraph{Heuristic for crypto API coverage}\label{subsec:analysisCapabilities}
\topicsentence{During our preliminary study, we identified that understanding the covered API classes is of importance for crypto-API misuse detectors.}
A detector which misses an API class may miss a vulnerability in an application while having a high precision and recall. 
Thus, for the assessment of the detectors, being aware of the covered classes is of importance. 
While test generators seem to be a good choice for this task and aid our aim of maintainability and extensibility of \obench\, they are inadequate.
Previous approaches summarized in a survey~\cite{mustafa2017comparative} have shown to be insufficient for generating and establishing a ground-truth needed for \obench.
Instead of using test generators, we propose employing a heuristic that harnesses the two other benchmarks of \obench\ to create an API coverage feature.

This feature is inspired by \hermes~\cite{reif2017hermes} which is a framework to assess collections like the \qualitas\ and uses queries to derive subsets suited for analyzing, e.g., a specific API.
With this heuristic, we can extract all classes and methods of the JCA and match them with our benchmarks for \emph{real-world applications} and \emph{analysis capabilities}.
We plan to use the heuristic to guide the effective generation of our benchmark. 
First, the heuristic will enable us to select a diverse set of real-world JCA usages to enrich them with the metadata.
Second, as we will develop the synthetic test cases ourselves, the heuristic allows us to include classes and methods for the JCA that are not covered by the \emph{real-world applications benchmark}. 
Hence, this heuristic approach covers at least all real-world parts of the API from the \emph{real-world applications benchmark} as well as additional classes and methods that only occur in the synthetic \emph{analysis capabilities benchmark}.
Overall, we assume that this heuristic approach should yield an almost complete coverage of JCA class and method usages for a highly used API like the JCA.

\subsection{Publishing and Deployment Plan}\label{subsec:coverageAPI}\label{subsec:publishingDeployment}
\topicsentence{Open access and ease of use are key factors for establishing a benchmark as describe in Section~\ref{sec:methodology}.}
Hence, it is important to publish \obench\ on GitHub with proper instructions and documentation. 
\authorResponse{To encourage, include, and harness community feedback, we will make the GitHub repository accessable during the creation of \obench, e.g., via pull requests.
We will also contact the authors of existing crypto API misuse detection tools and benchmarks (cf.~Table~\ref{tab:benchToSA}).}

Here we present our plans \authorResponse{for the GitHub repository.}\
We will provide execution support for \obench\ similar to \mubench~\cite{amann2016mubench}.
Thus, alongside the benchmark itself, we will provide the tooling we use for the generation together with documentation to support updating, maintaining, extending, and contributing to \obench.
In addition, we plan to improve the accessibility of our tool by providing essential information, such as the number of test cases or the covered API classes, directly on our \textit{README}-page and using the GitHub release functionality. 
To ensure the continuous quality of our benchmark, we will use GitHub actions for continious deployment.
Furthermore, \obench\ will be open access with a unique Digital Object Identifier (DOI).

\subsection{Evaluation plan}\label{subsec:evelauation}
After creating the first version of \obench\ we will perform the evaluation to answer \hyperref[RQ_Evaluation]{\textbf{RQ3}}.
The evaluation will consist of two main aspects:
First, how do current crypto API misuse detection tools perform on \obench?
The primary motivation for this registered report is comparing crypto API misuse detection tools with a fair benchmark explicitly developed for this domain.
Once \obench\ is created, such a comparison and evaluation will be possible and therefore will be conducted.
\authorResponse{Futhermore, we envision to provide a score section, similar to OWASP~\cite{wichers_owasp_nodate}, in the GitHub repository of \obench\ where researchers can add their results with experiment docmentation via pull request.}
Second, we want to know how current crypto API misuse detection tool benchmarks compare to \obench\ in practice.
\authorResponse{We expect this to be a theoretical comparison as the approaches and goals of existing benchmarks are diverse (cf.~Table~\ref{tab:benchToSA}).}
The generation approach of \obench\ differs from existing benchmarks (cf. Section~\ref{sec:background}, Table~\ref{tab:benchToSA}) and is intended to serve as a reference benchmark for the community. 
Therefore, we expect \obench\ to differ from the other custom benchmarks in its composition and evaluation of the analyses. 
Hence, we will investigate these differences, also to understand how our generation process for the test cases compares against the instantiated benchmarks. 
Additionally, we will evaluate our heuristic approach for measuring the coverage of the JCA.

%% file: sections/70_validity.tex
\section{Threats to validity}
\label{sec:validity}
For the first version of \obench, we introduced some limitations (cf. Section~\ref{sec:intro}) to have a clear target domain and develop a tangible benchmark.
These limitations are choosing a specific programing language in Java, a specific API in the JCA, and finally focusing on static analysis tools.
However, our methodology (cf. Section~\ref{sec:methodology}) is designed for semi-automated benchmark generation and extensibility.
Hence, extending \obench\ to support other APIs or dynamic analyses has very much been taken into consideration during the initial development as a reasonable future step and is as a result not inhibited whatsoever by the design.
Furthermore, instantiating benchmarks for other programming languages by applying our methodology is also possible.
However, these extensions would each require a lot of additional effort and leave a benchmark that is significantly changed in shape and size
as the new real-world applications need to be mined with new criteria, and thus, the benchmarks for analysis capabilities and API coverage need adaptation, too.

Besides this, the generation of the benchmarks for \emph{analysis capabilities} and \emph{API coverage} are subject to manual work by the authors.
Since the authors are experts regarding static analysis and crypto API misuse detection, the chosen domain limitations make them an appropriate choice to develop and label the test cases. 
Furthermore, community feedback is part of the generation process and will therefore help to mitigate this threat to validity.

Lastly, employing the heuristic for the \emph{API coverage} benchmark is a new approach that we will evaluate.
Employing the heuristic seems to be a reasonable and practical approach to create the \emph{API coverage feature}.
However, this approach is untested and needs evaluation.

%% file: sections/60_Implications.tex
\section{Implications}
\label{sec:implications}
Benchmarks generally drive development and research in their fields as they provide measurable optimization goals~\cite{DaCapo06}. 
Therefore, establishing a reference benchmark for the domain of benchmarking crypto API misuse detection tools in the form of \obench\ with its open and transparent generation process will help the community manyfold:
\obench\ will allow to evaluate and compare current crypto API misuse detection tools on the same benchmark which has not been done, yet.
Having this comparison, future development can reference the comparison as well as \obench\ in new contributions.

Furthermore, by deriving, extending, and defining the process for the creation of \obench, we hope to provide a template for creating and maintaining benchmarks for specific domains in need of a reference benchmark and thereby steering current developments towards more open and reproducible research.